\documentclass[sigconf]{acmart}
\AtBeginDocument{%
  }

\setcopyright{acmlicensed}
\definecolor{jsonkey}{rgb}{0.6, 0.2, 0.2}
\definecolor{jsonstring}{rgb}{0.2, 0.6, 0.2}

\definecolor{BlueBG}{rgb}{0,0.46,0.71}

\definecolor{innerboxcolor}{HTML}{F0A35D} 
\definecolor{outerboxcolor}{HTML}{FAE2CA} 
\definecolor{lightblue}{HTML}{FAE2CA}

\usepackage{tikz}
\usetikzlibrary{decorations.pathmorphing}

\definecolor{upblue}{HTML}{388E3C}
\definecolor{downred}{HTML}{E53935}

\definecolor{darkgreen}{RGB}{0,128,0}  
\definecolor{darkred}{RGB}{139,0,0}  

\definecolor{myblue}{RGB}{31,119,180}    
\definecolor{mygreen}{RGB}{44,160,44}    
\definecolor{myorange}{RGB}{255,127,14}  
\definecolor{mypurple}{RGB}{153, 102, 204}  

\title{Urban-MAS: Human-Centered Urban Prediction with \\ LLM-Based Multi-Agent System}

\author{Shangyu Lou}
\email{shangyulou@ucsb.edu, slou4820@sdsu.edu}

\affiliation{%
  \institution{University of California, Santa Barbara \& San Diego State University}
  \state{California}
  \country{USA}
}

\begin{document}


\begin{abstract}



Urban Artificial Intelligence (Urban AI) has advanced human-centered urban tasks such as perception prediction and human dynamics. Large Language Models (LLMs) can integrate multimodal inputs to address heterogeneous data in complex urban systems but often underperform on domain-specific tasks. Urban-MAS, an LLM-based Multi-Agent System (MAS) framework, is introduced for human-centered urban prediction under zero-shot settings. It includes three agent types: Predictive Factor Guidance Agents, which prioritize key predictive factors to guide knowledge extraction and enhance the effectiveness of compressed urban knowledge in LLMs; Reliable UrbanInfo Extraction Agents, which improve robustness by comparing multiple outputs, validating consistency, and re-extracting when conflicts occur; and Multi-UrbanInfo Inference Agents, which integrate extracted multi-source information across dimensions for prediction. Experiments on running-amount prediction and urban perception across Tokyo, Milan, and Seattle demonstrate that Urban-MAS substantially reduces errors compared to single-LLM baselines. Ablation studies indicate that Predictive Factor Guidance Agents are most critical for enhancing predictive performance, positioning Urban-MAS as a scalable paradigm for human-centered urban AI prediction. Code is available on the project website.\footnote{\url{https://github.com/THETUREHOOHA/UrbanMAS}}

\end{abstract}

\ccsdesc[500]{Computing methodologies~Spatial and physical reasoning}

\copyrightyear{2025}
\acmYear{2025}
\setcopyright{cc}
\setcctype{by}
\acmConference[UrbanAI '25]{The 3rd ACM SIGSPATIAL International Workshop
on Advances in Urban-AI}{November 3--6, 2025}{Minneapolis, MN, USA}
\acmBooktitle{The 3rd ACM SIGSPATIAL International Workshop on Advances in Urban-AI (UrbanAI '25), November 3--6, 2025, Minneapolis, MN, USA}
\acmDOI{10.1145/3764926.3771951}
\acmISBN{979-8-4007-2189-2/2025/11}
\keywords{Large Language Models, Urban AI,  Multi-Agent System, Human-centered Urban Prediction}



\maketitle

\section{Introduction}
Urban AI has played an increasingly important role in addressing diverse urban challenges. Human-centered studies—such as urban perception~\cite{Zhang2018MeasuringHumanPerceptions,lou2024assessing,11118366} and human dynamics~\cite{Dong2023}, deepening understanding of urban systems and support evidence-based policymaking to improve quality of life~\cite{Zhang2018MeasuringHumanPerceptions,lou2024assessing,Dong2023}. However, the complexity of urban systems makes selecting and representing predictive features difficult, limiting accuracy~\cite{Feng2024UrbanLLaVA}. Large Language Models (LLMs)\footnote{Also referred to as MLLMs; used interchangeably in this paper.} have shown strong potential in integrating heterogeneous modalities like text and imagery~\cite{Feng2024UrbanLLaVA}. However, it remains limited in domain-specific applications~\cite{Zhang2023SurveyLLM,ke2025demystifyingdomainadaptiveposttrainingfinancial,ke2025naacl2025tutorialadaptationlarge,ke2023continualpretraininglanguagemodels,ke2023continuallearningnaturallanguage}. LLMs often compress vast knowledge, making it difficult to model complex domain problems~\cite{Manvi2023GeoLLM,10800533}, and single-LLM approaches struggle to handle the specialized and multifaceted requirements of urban tasks~\cite{Kalyuzhnaya2025}, leading to biased or incomplete outputs that may misinform policymaking. To overcome these limitations, researchers are increasingly adopting MAS~\cite{Ke2025LLMReasoning,Ke2025MASZero}, where multiple LLM-based agents collaborate. Compared with single LLMs, MAS offers stronger specialization and fault tolerance, mitigating common issues such as hallucinations and insufficient domain expertise~\cite{Kalyuzhnaya2025}. Through division of labor and collaborative reasoning, MAS demonstrates improved scalability and reasoning ability for complex urban tasks~\cite{Ke2025LLMReasoning}. However, to the best of knowledge, no prior work has applied MAS approaches to enhance human-centered urban predictions, motivating this work to examine: To what extent do LLM-based multi-agent systems enhance human-centered urban prediction?

Existing approaches to enhance single-LLM performance mainly rely on fine-tuning~\cite{Feng2024UrbanLLaVA,Manvi2023GeoLLM,Zeng2025FineEdit}, which demands extensive data and computation, or on chain-of-thought (CoT) prompting~\cite{Li2024}, which requires manually designed reasoning chains that are time-consuming and labor-intensive. In human-centered urban prediction such as urban perception prediction, much attention has been given to identifying the most important features that contribute to task accuracy~\cite{Liu2023}. yet none explicitly extract the most predictive influential factors to guide LLMs. Given the complexity of cities, determinants span multiple dimensions (social and built environments) and scales (macro and street levels)~\cite{Dong2023}, making manual identification highly demanding. By contrast, MAS enables parallelized and specialized exploration\cite{Anthropic2024} across dimensions, providing a more automatic, comprehensive, and efficient way to uncover task-relevant influential factors. While deep-research MAS has shown promise in general AI~\cite{Anthropic2024}, it has not been applied to systematically identify such factors in human-centered urban prediction.

In addition, Some studies attempt to leverage task-related knowledge compressed within a single LLM~\cite{Li2024,Manvi2023GeoLLM}. However, they lack mechanisms to verify and enhance the reliability of extracted information. Biased or inconsistent information, once incorporated into prediction, can cause errors, thereby reduce result reliability. MAS offers the advantages through role allocation and task specialization\cite{Anthropic2024,Ke2025LLMReasoning}, it could enables simultaneous cross-validation and refinement during knowledge extraction, substantially improving credibility~\cite{Kalyuzhnaya2025}. Nevertheless, this capability has yet to be applied to human-centered urban tasks.

To address these limitations and explore MAS potential in urban prediction, this study introduces Urban-MAS, a multi-agent system for human-centered urban tasks under zero-shot conditions. Urban-MAS integrates three classes of agents: (1) Predictive Factor Guidance Agents, which collaborate to prioritize influential factors and guide knowledge extraction; (2) Reliable UrbanInfo Extraction Agents, which improve robustness through multi-output comparison, consistency validation, and re-extraction when conflicts occur; and (3) Multi-UrbanInfo Inference Agents, which integrate multi-source information for prediction. This framework overcomes the constraints of single-LLM methods and advances LLM performance in human-centered urban prediction. The main contributions are:
\begin{itemize}
\item Presents the first MAS framework for human-centered urban prediction. Urban-MAS integrates three agent layers to enable prioritized factor selection, reliable urban information extraction, and integrated multi-source prediction under zero-shot conditions, outperforming single-LLM baselines across metrics.  

\item Introduces a MAS-based mechanism to improve the reliability of urban knowledge extraction. Reliable UrbanInfo Extraction Agents use output comparison, consistency checks, and selective re-extraction to mitigate bias and enhance predictive robustness.  

\item Evaluates performance on urban perception and human dynamics prediction across cities on three continents. Results demonstrate that Urban-MAS achieves efficient, low-cost, and significant zero-shot gains, advancing human-centered Urban AI research.  
\end{itemize}


\section{Urban-MAS}

 \begin{figure*}[h]
  \centering
  \includegraphics[width=\linewidth]{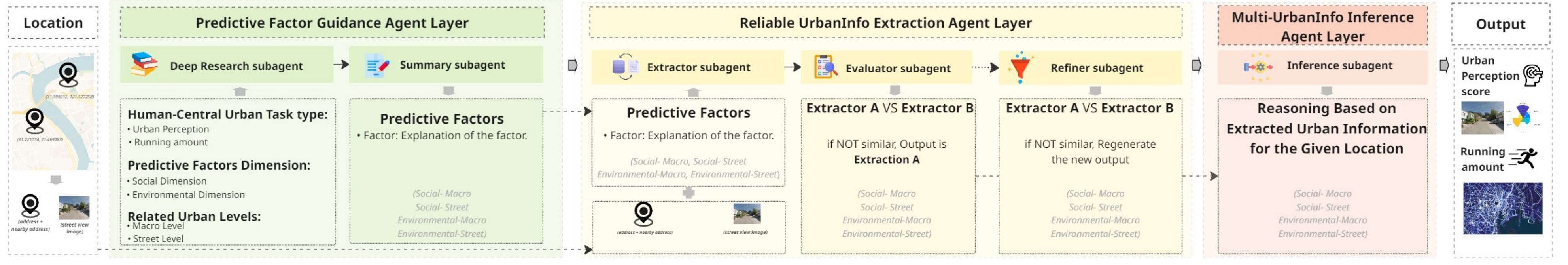}
 \vspace{-\baselineskip}
  \caption{\small Urban-MAS comprises three agent layers: (Left) Predictive Factor Guidance Agents prioritize factor selection; (Middle) Reliable UrbanInfo Extraction Agents ensure reliable urban information extraction through consistency checks; (Right) Multi-UrbanInfo Inference Agents integrate multi-source urban information for robust urban prediction.}
  \Description{A woman and a girl in white dresses sit in an open car.}
\end{figure*}
To enhance LLM performance in human-centered urban research, the Urban-MAS framework comprises three agent layers (Figure 1). Predictive Factor Guidance Agents identify the most relevant predictive factors, guiding extraction and improving the utility of compressed urban knowledge. Reliable UrbanInfo Extraction Agents enhance stability by generating multiple outputs, checking consistency, and re-extracting when needed to ensure trustworthy information. Multi-UrbanInfo Inference Agents integrate these refined multi-source signals across dimensions and scales to deliver robust, task-specific urban predictions.\\
\textbf{Predictive Factor Guidance Agents.} Urban prediction requires focusing on the most influential factors for each task and data source, as general prompts often yield noisy cues. To address this, the Predictive Factor Guidance Agents layer employs targeted \textit{Deep-research} subagents based on the opendeepresearch framework from LangChain\cite{LangChain2024OpenDeepResearch} to generate research-level reports on the most influential factors of the task. Then, the \textit{Summary} subagents summarize the findings into concise predictive factors, organized by social and environmental dimensions and macro and street levels. Given an urban prediction task $\tau$, the input is the task description. Dimensions are defined as $D = {\text{Social}, \text{Built Environmental}}$ and levels as $R = {\text{Macro}, \text{Street}}$. For each $(d, r)$ pair under the task, deep-research subagents produce a brief report $t_{d,r}$ containing six key predictive factors, which summary subagents compress into a predictive factor set $P_{d,r}$ as output, providing dimension- and level-specific guidance for the UrbanInfo Extraction Agents.\\
\textbf{Reliable UrbanInfo Extraction Agents.}
Urban information from a single LLM call is often noisy or inconsistent, weakening inference. To enhance stability, each extraction agent generates two output variants via an \textit{Extractor} subagent and compares them with an \textit{Evaluator} subagent. Conflicting fields are selectively re-extracted by a \textit{Refiner} subagent. For each location $\ell$, four extraction agents are aligned to $P_{d,r}$: social-macro, social-street, environment-macro, and environment-street. Candidate outputs are compared by similarity metrics; if agreement exceeds a threshold, one is accepted, otherwise only conflicting fields are regenerated, yielding reliable outputs $U_{d,r}$. This dual-variant and conflict-repair mechanism ensures consistent UrbanInfo across dimensions and levels, strengthening multi-source integration. Specifically, for each dimension-scale pair, the \textit{Evaluator} subagent compares two urban information independently generated variants (A and B) from \textit{Extractor}. The two generated variants are normalized (lowercasing, punctuation removal, whitespace collapsing), and field-level hybrid soft similarity is computed as
\begin{equation}
\small
\texttt{soft\_sim}(a,b) = 0.4 \times \text{Jaccard}(a,b) + 0.6 \times \text{SequenceMatcher}(a,b),
\end{equation}
where \textit{Jaccard}\cite{Travieso2024Jaccard} measures token overlap and \textit{SequenceMatcher}\cite{PythonDifflib} captures phrase-level alignment. A stability threshold of $0.72$ is used: if similarity $\geq 0.72$, Variant~A is accepted; otherwise, the \textit{Refiner} regenerates only differing fields.\\
\textbf{Multi-UrbanInfo Inference Agents.}
Urban prediction requires reasoning over complementary UrbanInfo across social and built dimensions and macro- and street-level scales. Even when each source is reliable, isolated reasoning risks imbalance. To address this, the inference layer employs an \textit{LLM-based Inference Agent} that jointly processes the four reliable inputs---$U^{*}_{\text{social,macro}}$, $U^{*}_{\text{social,street}}$, $U^{*}_{\text{environment,macro}}$, and $U^{*}_{\text{environment,street}}$---to infer task-specific outputs such as running amount or perception scores. The Inference Agent receives the four structured JSON inputs, enforces schema constraints (e.g., \texttt{\{"running\_amount": 0.0\}}). As the final Urban-MAS stage, this layer integrates reliable extraction to deliver robust, coherent predictions.

\section{Experiments}
\textbf{Tasks.}
We evaluate Urban-MAS on two representative human-centered urban prediction tasks: (i) urban perception prediction, focusing on one positive (lively) and one negative (boring) dimension of perception, and (ii) human dynamics prediction via running amount estimation. These tasks capture complementary aspects of human–environment interaction by linking perceptual and behavioral responses to urban form.\\
\textbf{Datasets.}
Experiments are conducted on 300 samples across Tokyo, Milan, and Seattle, covering three continents to assess cross-regional generalizability.For each sampled location, we process the raw geographic coordinates into location text via OpenStreetMap’s Nominatim API for reverse geocoding (convert coordinates to address), and further enrich the inputs by querying nearby points of interest using the Overpass API. In addition, street-view imagery is retrieved through the Google Maps API. These inputs are used as multi-source data for Urban-MAS across different human-centered tasks. Although the validation dataset is relatively small, its scale is constrained by the cost of collecting and verifying multi-source urban data (street-view imagery, POI queries), yet remains sufficient for evaluating the quality of the LLM-based solution using several hundred representative examples\cite{Kalyuzhnaya2025}.\\
For running amount prediction, we use Strava heatmaps , a widely recognized source of physical activity data. Following Le’s agent-based modeling approach, high-resolution Strava heatmaps were collected for each study area, and raster values were extracted in QGIS at each sampling point. Brightness values, representing running intensity, were rescaled to [0,10]. 
For urban perception, we adopt Place Pulse 2.0 \cite{Zhang2018MeasuringHumanPerceptions}, the largest benchmark of human perceptions of urban environments. Pairwise Google Street View comparisons were aggregated using the TrueSkill algorithm, yielding continuous perception scores that we rescaled to [0,10]. \\
\textbf{Models.}
The experimental setup was designed to evaluate the performance and capabilities of LLM-based multi-agent systems in human-centered urban perception and prediction. All experiments were conducted using GPT-5, a closed-source model representing the state of the art in its class. The exclusive use of LLM-based configurations aims to examine how multi-agent interaction improves efficiency, reasoning, and overall performance compared with a single-LLM baseline, as well as to assess the contribution of each agent component within the MAS framework. The proposed Urban-MAS framework comprises three layers of agents: Predictive Factor Guidance Agents, Reliable UrbanInfo Extraction Agents, and Multi-UrbanInfo Inference Agents. The Predictive Factor Guidance layer leverages the opendeepresearch framework from LangChain—ranked sixth on the Deep Research Bench Leaderboard with GPT-4o—to ensure consistency and relevance of the extracted predictive factors. The Reliable UrbanInfo Extraction and Multi-UrbanInfo Inference layers are implemented on GPT-5 in JSON mode using zero-shot prompting, with a single-LLM GPT-5 baseline included for comparison.\\
\textbf{Evaluation.}
Performance is assessed using MAE, MSE, and RMSE for continuous outcomes (running amount and perception scores). We also report ablation studies to isolate the contributions of factor guidance and reliability enhancement.

\subsection{Main Results}
\begin{table}[t]
  \caption{\small Performance gains with Urban-MAS with single LLM across multiple tasks (all metrics: lower is better).}
   \vspace{-\baselineskip}
  \label{tab:comparison_metrics_vertical}
  \centering
  \small
  \begin{tabular}{lccc}
    \toprule
    Method & MAE & MSE & RMSE \\
    \midrule
    \multicolumn{4}{l}{\textit{People Running Amount}} \\
    Single LLM & 2.99 & 13.73 & 3.70 \\
    Urban-MAS (Ours) & \textbf{2.97} (↓0.73\%) & \textbf{13.20} (↓3.82\%) & \textbf{3.63} (↓1.93\%) \\
    \midrule
    \multicolumn{4}{l}{\textit{People Urban Perception (Boringness)}} \\
    Single LLM & 2.83 & 9.95 & 3.15 \\
    Urban-MAS (Ours) & \textbf{2.05} (↓27.37\%) & \textbf{5.84} (↓41.33\%) & \textbf{2.42} (↓23.40\%) \\
    \midrule
    \multicolumn{4}{l}{\textit{People Urban Perception (Liveliness)}} \\
    Single LLM & 2.69 & 9.10 & 3.02 \\
    Urban-MAS (Ours) & \textbf{1.73} (↓35.81\%) & \textbf{4.40} (↓51.67\%) & \textbf{2.10} (↓30.48\%) \\
    \bottomrule
  \end{tabular}
\end{table}

Table 1 summarizes the performance improvements achieved by the integrated Urban-MAS compared to the baseline single-LLM model (GPT-5). Relative to GPT-5, Urban-MAS achieves substantial reductions across all error metrics for every task, demonstrating its effectiveness in enhancing prediction performance for human-centered urban applications. A closer look at the baseline results further reveals variations in the degree of improvement across task types: in particular, error reductions are more pronounced for the safety perception task than for the running amount prediction task.

\subsection{Ablation Study}

\begin{table}[h]
  \caption{\small Ablation study on Predictive Factor Guidance and Reliable UrbanInfo Extraction (all metrics: lower is better).}
     \vspace{-\baselineskip}
  \label{tab:main_metrics_vertical}
  \centering
  \small
  \setlength{\tabcolsep}{4pt}
  \begin{tabular}{lccc}
    \toprule
    Method Variant & MAE & MSE & RMSE \\
    \midrule
    \multicolumn{4}{l}{\textit{People Running Amount}} \\
    Urban-MAS & \textbf{2.97} & \textbf{13.20} & \textbf{3.63} \\
    \quad - PredictiveFactors & 4.53 (\(\uparrow\)52.84\%) & 26.80 (\(\uparrow\)102.98\%) & 5.18 (\(\uparrow\)42.47\%) \\
    \quad - ReliabilityBoost & 2.98 (\(\uparrow\)0.30\%) & 13.39 (\(\uparrow\)1.46\%) & 3.66 (\(\uparrow\)0.73\%) \\
    \midrule
    \multicolumn{4}{l}{\textit{People Urban Perception (Boringness)}} \\
    Urban-MAS & \textbf{2.05} & \textbf{5.84} & \textbf{2.42} \\
    \quad - PredictiveFactors & 2.39 (\(\uparrow\)16.37\%) & 7.52 (\(\uparrow\)28.69\%) & 2.74 (\(\uparrow\)13.44\%) \\
    \quad - ReliabilityBoost & 2.29 (\(\uparrow\)11.52\%) & 6.98 (\(\uparrow\)19.45\%) & 2.64 (\(\uparrow\)9.29\%) \\
    \midrule
    \multicolumn{4}{l}{\textit{People Urban Perception (Liveliness)}} \\
    Urban-MAS & \textbf{1.73} & \textbf{4.40} & \textbf{2.10} \\
    \quad - PredictiveFactors & 2.54 (\(\uparrow\)46.89\%) & 8.09 (\(\uparrow\)83.79\%) & 2.84 (\(\uparrow\)35.57\%) \\
    \quad - ReliabilityBoost & 2.21 (\(\uparrow\)28.06\%) & 6.47 (\(\uparrow\)47.02\%) & 2.54 (\(\uparrow\)21.25\%) \\
    \bottomrule
  \end{tabular}
\end{table}

We conducted an ablation study to evaluate two strategies for improving LLM performance on human-centered urban tasks: (1) Predictive Factor Guidance Agents, which prioritize influential predictive factors, and (2) Reliable UrbanInfo Extraction Agents, which enhance extraction reliability. As shown in Table 2, the full Urban-MAS configuration achieves the lowest errors across all tasks. Removing the Predictive Factor Guidance module causes a larger error increase than removing the Reliability module, indicating that prioritizing predictive factors is more critical .Removing the Predictive Factor Guidance module causes a larger error increase than removing the Reliability module, indicating that prioritizing predictive factors is more critical. Disabling factor prioritization (“–PredictiveFactors”) consistently raises errors, especially in running amount prediction, showing its importance for modeling activity patterns. Conversely, disabling reliability enhancement (“–ReliabilityBoost”) increases errors across all tasks, particularly in perception-related dimensions such as liveliness and boringness, highlighting the necessity of reliable urban information extraction.

\subsection{Case Study}
\textbf{Predictive Factors Agent.}
Predictive factors were derived through an LLM-guided process across two dimensions (social and environmental) and two spatial scales (macro and street level). These structured factor sets serve as measurable descriptors guiding information extraction and inference within Urban-MAS. The complete list of predictors are available on the project website.\\
\textbf{UrbanInfo Extraction Agents.}
The Reliable UrbanInfo Extraction Agents Layers extract consistent urban information for each location. When both variants generated from Extractor subagent, the Evaluator subagent confirms stability and retains one version directly. However, when factual discrepancies arise, the Refiner subagent performs conflict-only reconciliation—regenerating only inconsistent fields while preserving verified ones. This selective correction ensures factual reliability without redundant regeneration. Detailed examples are available on the project website.

\section{Conclusion} 
This paper introduced Urban-MAS, a novel LLM-based multi-agent system for human-centered urban tasks with multi-source data inputs. Urban-MAS integrates automated prioritization of the most predictive factors to guide subsequent knowledge extraction, enhanced reliability in extracting urban information from multi-source data, and predictor agents that collaborate under a zero-shot setting to improve performance on human-centered urban tasks. Experimental results across multiple cities and tasks demonstrate that, compared with single-LLM baselines, Urban-MAS significantly reduces prediction error. Urban-MAS also provides methodological insights: prioritizing the most predictive factors is crucial for enhancing human-centered urban prediction tasks, and improving the reliability of extracting urban knowledge from LLMs is also essential. Future work will extend this framework by incorporating MAS-based automatic optimization of urban prediction performance and applying Urban-MAS to a broader range of urban tasks and larger test samples.


\bibliographystyle{ACM-Reference-Format}
\bibliography{sample-base2,zixuan}


\begin{thebibliography}{22}


\ifx \showCODEN    \undefined \def \showCODEN     #1{\unskip}     \fi
\ifx \showISBNx    \undefined \def \showISBNx     #1{\unskip}     \fi
\ifx \showISBNxiii \undefined \def \showISBNxiii  #1{\unskip}     \fi
\ifx \showISSN     \undefined \def \showISSN      #1{\unskip}     \fi
\ifx \showLCCN     \undefined \def \showLCCN      #1{\unskip}     \fi
\ifx \shownote     \undefined \def \shownote      #1{#1}          \fi
\ifx \showarticletitle \undefined \def \showarticletitle #1{#1}   \fi
\ifx \showURL      \undefined \def \showURL       {\relax}        \fi
\providecommand\bibfield[2]{#2}
\providecommand\bibinfo[2]{#2}
\providecommand\natexlab[1]{#1}
\providecommand\showeprint[2][]{arXiv:#2}

\bibitem[{Anthropic}(2024)]%
        {Anthropic2024}
\bibfield{author}{\bibinfo{person}{{Anthropic}}.} \bibinfo{year}{2024}\natexlab{}.
\newblock \bibinfo{title}{How we built our multi-agent research system}.
\newblock \bibinfo{howpublished}{\url{https://www.anthropic.com/engineering/multi-agent-research-system}}.
\newblock


\bibitem[Cui and Lou(2025)]%
        {11118366}
\bibfield{author}{\bibinfo{person}{Ziqi Cui} {and} \bibinfo{person}{Shangyu Lou}.} \bibinfo{year}{2025}\natexlab{}.
\newblock \showarticletitle{Syncperception: A Real-Time Urban Perception Prediction Tool Based on Graph Neural Networks}. In \bibinfo{booktitle}{\emph{2025 Annual Modeling and Simulation Conference (ANNSIM)}}.
\newblock


\bibitem[Dong et~al\mbox{.}(2023)]%
        {Dong2023}
\bibfield{author}{\bibinfo{person}{Lin Dong}, \bibinfo{person}{Hongchao Jiang}, \bibinfo{person}{Wenjing Li}, \bibinfo{person}{Bing Qiu}, \bibinfo{person}{Hao Wang}, {and} \bibinfo{person}{Waishan Qiu}.} \bibinfo{year}{2023}\natexlab{}.
\newblock \showarticletitle{Assessing impacts of objective features and subjective perceptions of street environment on running amount: A case study of Boston}.
\newblock \bibinfo{journal}{\emph{Landscape and Urban Planning}}  \bibinfo{volume}{235} (\bibinfo{year}{2023}), \bibinfo{pages}{104756}.
\newblock


\bibitem[Feng et~al\mbox{.}(2024)]%
        {Feng2024UrbanLLaVA}
\bibfield{author}{\bibinfo{person}{Jie Feng}, \bibinfo{person}{Shengyuan Wang}, \bibinfo{person}{Tianhui Liu}, \bibinfo{person}{Yanxin Xi}, {and} \bibinfo{person}{Yong Li}.} \bibinfo{year}{2024}\natexlab{}.
\newblock \showarticletitle{UrbanLLaVA: A Multi-modal Large Language Model for Urban Intelligence with Spatial Reasoning and Understanding}.
\newblock \bibinfo{journal}{\emph{arXiv preprint}} (\bibinfo{year}{2024}).
\newblock
\showeprint[arxiv]{2406.05294}


\bibitem[Kalyuzhnaya et~al\mbox{.}(2025)]%
        {Kalyuzhnaya2025}
\bibfield{author}{\bibinfo{person}{Anna Kalyuzhnaya}, \bibinfo{person}{Sergey Mityagin}, \bibinfo{person}{Elizaveta Lutsenko}, \bibinfo{person}{Andrey Getmanov}, \bibinfo{person}{Yaroslav Aksenkin}, \bibinfo{person}{Kamil Fatkhiev}, \bibinfo{person}{Kirill Fedorin}, \bibinfo{person}{Nikolay~O. Nikitin}, \bibinfo{person}{Natalia Chichkova}, \bibinfo{person}{Vladimir Vorona}, {and} \bibinfo{person}{Alexander Boukhanovsky}.} \bibinfo{year}{2025}\natexlab{}.
\newblock \showarticletitle{LLM Agents for Smart City Management: Enhancing Decision Support Through Multi-Agent AI Systems}.
\newblock \bibinfo{journal}{\emph{Smart Cities}} \bibinfo{volume}{8}, \bibinfo{number}{1} (\bibinfo{year}{2025}), \bibinfo{pages}{19}.
\newblock


\bibitem[Ke et~al\mbox{.}(2025a)]%
        {Ke2025LLMReasoning}
\bibfield{author}{\bibinfo{person}{Zixuan Ke}, \bibinfo{person}{Fangkai Jiao}, \bibinfo{person}{Yifei Ming}, \bibinfo{person}{Xuan-Phi Nguyen}, \bibinfo{person}{Austin Xu}, \bibinfo{person}{Do~Xuan Long}, \bibinfo{person}{Minzhi Li}, \bibinfo{person}{Chengwei Qin}, \bibinfo{person}{Peifeng Wang}, \bibinfo{person}{Silvio Savarese}, \bibinfo{person}{Caiming Xiong}, {and} \bibinfo{person}{Shafiq Joty}.} \bibinfo{year}{2025}\natexlab{a}.
\newblock \showarticletitle{A Survey of Frontiers in {LLM} Reasoning: Inference Scaling, Learning to Reason, and Agentic Systems}.
\newblock \bibinfo{journal}{\emph{arXiv preprint arXiv:2504.09037}} (\bibinfo{year}{2025}).
\newblock


\bibitem[Ke and Liu(2023)]%
        {ke2023continuallearningnaturallanguage}
\bibfield{author}{\bibinfo{person}{Zixuan Ke} {and} \bibinfo{person}{Bing Liu}.} \bibinfo{year}{2023}\natexlab{}.
\newblock \bibinfo{title}{Continual Learning of Natural Language Processing Tasks: A Survey}.
\newblock
\showeprint[arxiv]{2211.12701}


\bibitem[Ke et~al\mbox{.}(2025b)]%
        {ke2025naacl2025tutorialadaptationlarge}
\bibfield{author}{\bibinfo{person}{Zixuan Ke}, \bibinfo{person}{Yifei Ming}, {and} \bibinfo{person}{Shafiq Joty}.} \bibinfo{year}{2025}\natexlab{b}.
\newblock \bibinfo{title}{NAACL2025 Tutorial: Adaptation of Large Language Models}.
\newblock
\showeprint[arxiv]{2504.03931}~[cs.CL]


\bibitem[Ke et~al\mbox{.}(2025c)]%
        {ke2025demystifyingdomainadaptiveposttrainingfinancial}
\bibfield{author}{\bibinfo{person}{Zixuan Ke}, \bibinfo{person}{Yifei Ming}, \bibinfo{person}{Xuan-Phi Nguyen}, \bibinfo{person}{Caiming Xiong}, {and} \bibinfo{person}{Shafiq Joty}.} \bibinfo{year}{2025}\natexlab{c}.
\newblock \bibinfo{title}{Demystifying Domain-adaptive Post-training for Financial LLMs}.
\newblock
\showeprint[arxiv]{2501.04961}


\bibitem[Ke et~al\mbox{.}(2023)]%
        {ke2023continualpretraininglanguagemodels}
\bibfield{author}{\bibinfo{person}{Zixuan Ke}, \bibinfo{person}{Yijia Shao}, \bibinfo{person}{Haowei Lin}, \bibinfo{person}{Tatsuya Konishi}, \bibinfo{person}{Gyuhak Kim}, {and} \bibinfo{person}{Bing Liu}.} \bibinfo{year}{2023}\natexlab{}.
\newblock \bibinfo{title}{Continual Pre-training of Language Models}.
\newblock
\showeprint[arxiv]{2302.03241}


\bibitem[Ke et~al\mbox{.}(2025d)]%
        {Ke2025MASZero}
\bibfield{author}{\bibinfo{person}{Zixuan Ke}, \bibinfo{person}{Austin Xu}, \bibinfo{person}{Yifei Ming}, \bibinfo{person}{Xuan-Phi Nguyen}, \bibinfo{person}{Caiming Xiong}, {and} \bibinfo{person}{Shafiq Joty}.} \bibinfo{year}{2025}\natexlab{d}.
\newblock \showarticletitle{{MAS-ZERO}: Designing Multi-Agent Systems with Zero Supervision}.
\newblock \bibinfo{journal}{\emph{arXiv preprint arXiv:2505.14996}} (\bibinfo{year}{2025}).
\newblock


\bibitem[{LangChain}(2024)]%
        {LangChain2024OpenDeepResearch}
\bibfield{author}{\bibinfo{person}{{LangChain}}.} \bibinfo{year}{2024}\natexlab{}.
\newblock \bibinfo{title}{Open Deep Research}.
\newblock \bibinfo{howpublished}{\url{https://blog.langchain.com/open-deep-research/}}.
\newblock


\bibitem[Li et~al\mbox{.}(2024a)]%
        {10800533}
\bibfield{author}{\bibinfo{person}{Xinjin Li}, \bibinfo{person}{Yu Ma}, \bibinfo{person}{Yangchen Huang}, \bibinfo{person}{Xingqi Wang}, \bibinfo{person}{Yuzhen Lin}, {and} \bibinfo{person}{Chenxi Zhang}.} \bibinfo{year}{2024}\natexlab{a}.
\newblock \showarticletitle{Synergized Data Efficiency and Compression (SEC) Optimization for Large Language Models}. In \bibinfo{booktitle}{\emph{2024 4th International Conference on Electronic Information Engineering and Computer Science (EIECS)}}.
\newblock


\bibitem[Li et~al\mbox{.}(2024b)]%
        {Li2024}
\bibfield{author}{\bibinfo{person}{Zongrong Li}, \bibinfo{person}{Junhao Xu}, \bibinfo{person}{Siqin Wang}, \bibinfo{person}{Yifan Wu}, {and} \bibinfo{person}{Haiyang Li}.} \bibinfo{year}{2024}\natexlab{b}.
\newblock \showarticletitle{StreetViewLLM: Extracting Geographic Information Using a Chain-of-Thought Multimodal Large Language Model}.
\newblock \bibinfo{journal}{\emph{arXiv preprint arXiv:2411.14476}} (\bibinfo{year}{2024}).
\newblock


\bibitem[Liu et~al\mbox{.}(2023)]%
        {Liu2023}
\bibfield{author}{\bibinfo{person}{Yunzhe Liu}, \bibinfo{person}{Meixu Chen}, \bibinfo{person}{Meihui Wang}, \bibinfo{person}{Jing Huang}, \bibinfo{person}{Fisher Thomas}, \bibinfo{person}{Kazem Rahimi}, {and} \bibinfo{person}{Mohammad Mamouei}.} \bibinfo{year}{2023}\natexlab{}.
\newblock \showarticletitle{An interpretable machine learning framework for measuring urban perceptions from panoramic street view images}.
\newblock \bibinfo{journal}{\emph{PLOS Computational Biology}} \bibinfo{volume}{19}, \bibinfo{number}{3} (\bibinfo{year}{2023}), \bibinfo{pages}{e1010911}.
\newblock


\bibitem[Lou et~al\mbox{.}(2024)]%
        {lou2024assessing}
\bibfield{author}{\bibinfo{person}{Shangyu Lou}, \bibinfo{person}{Gabriele Stancato}, {and} \bibinfo{person}{Barbara~EA Piga}.} \bibinfo{year}{2024}\natexlab{}.
\newblock \showarticletitle{Assessing in-motion urban visual perception: analyzing urban features, design qualities, and people’s perception}.
\newblock In \bibinfo{booktitle}{\emph{Advances in Representation: New AI-and XR-Driven Transdisciplinarity}}. \bibinfo{publisher}{Springer}, \bibinfo{pages}{691--706}.
\newblock


\bibitem[Manvi et~al\mbox{.}(2023)]%
        {Manvi2023GeoLLM}
\bibfield{author}{\bibinfo{person}{Rohin Manvi}, \bibinfo{person}{Samar Khanna}, \bibinfo{person}{Gengchen Mai}, \bibinfo{person}{Marshall Burke}, \bibinfo{person}{David~B. Lobell}, {and} \bibinfo{person}{Stefano Ermon}.} \bibinfo{year}{2023}\natexlab{}.
\newblock \showarticletitle{GeoLLM: Extracting Geospatial Knowledge from Large Language Models}.
\newblock \bibinfo{journal}{\emph{arXiv preprint}} (\bibinfo{year}{2023}).
\newblock
\showeprint[arxiv]{2310.06213}


\bibitem[{Python Library}({[n.\,d.]})]%
        {PythonDifflib}
\bibfield{author}{\bibinfo{person}{{Python Library}}.} \bibinfo{year}{[n.\,d.]}\natexlab{}.
\newblock \bibinfo{title}{Difflib Python Libaray}.
\newblock \bibinfo{howpublished}{\url{https://docs.python.org/3/library/difflib.html}}.
\newblock


\bibitem[Travieso et~al\mbox{.}(2024)]%
        {Travieso2024Jaccard}
\bibfield{author}{\bibinfo{person}{Gonzalo Travieso}, \bibinfo{person}{Alexandre Benatti}, {and} \bibinfo{person}{Luciano da F.~Costa}.} \bibinfo{year}{2024}\natexlab{}.
\newblock \showarticletitle{An Analytical Approach to the Jaccard Similarity Index}.
\newblock \bibinfo{journal}{\emph{arXiv preprint arXiv:2410.16436}} (\bibinfo{year}{2024}).
\newblock


\bibitem[Zeng et~al\mbox{.}(2025)]%
        {Zeng2025FineEdit}
\bibfield{author}{\bibinfo{person}{Yiming Zeng}, \bibinfo{person}{Wanhao Yu}, \bibinfo{person}{Zexin Li}, \bibinfo{person}{Tao Ren}, \bibinfo{person}{Yu Ma}, \bibinfo{person}{Jinghan Cao}, \bibinfo{person}{Xiyan Chen}, {and} \bibinfo{person}{Tingting Yu}.} \bibinfo{year}{2025}\natexlab{}.
\newblock \bibinfo{title}{Bridging the Editing Gap in LLMs: FineEdit for Precise and Targeted Text Modifications}.
\newblock
\showeprint[arxiv]{2502.13358}


\bibitem[Zhang et~al\mbox{.}(2018)]%
        {Zhang2018MeasuringHumanPerceptions}
\bibfield{author}{\bibinfo{person}{Fan Zhang}, \bibinfo{person}{Bolei Zhou}, \bibinfo{person}{Liu Liu}, \bibinfo{person}{Yu Liu}, \bibinfo{person}{Helene~H. Fung}, \bibinfo{person}{Hui Lin}, {and} \bibinfo{person}{Carlo Ratti}.} \bibinfo{year}{2018}\natexlab{}.
\newblock \showarticletitle{Measuring human perceptions of a large-scale urban region using machine learning}.
\newblock \bibinfo{journal}{\emph{Landscape and Urban Planning}}  \bibinfo{volume}{180} (\bibinfo{year}{2018}), \bibinfo{pages}{148--160}.
\newblock


\bibitem[Zhang et~al\mbox{.}(2023)]%
        {Zhang2023SurveyLLM}
\bibfield{author}{\bibinfo{person}{Sanqiang Zhang}, \bibinfo{person}{Xinyu Liu}, {et~al\mbox{.}}} \bibinfo{year}{2023}\natexlab{}.
\newblock \showarticletitle{A Survey on Large Language Models: Applications, Challenges, and Opportunities}.
\newblock \bibinfo{journal}{\emph{arXiv preprint arXiv:2305.18703}} (\bibinfo{year}{2023}).
\newblock


\end{thebibliography}

\end{document}